\documentclass{IEEEtran}
\usepackage{cite}
\usepackage{amsmath,amssymb,amsfonts}
\usepackage{algorithmic}
\usepackage{graphicx,subfigure}
\usepackage{textcomp}
\def\e{\begin{equation}}
\def\f{\end{equation}}
\def\l#1{\label{eq:#1}}
\def\r#1{(\ref{eq:#1})}
\def\BibTeX{{\rm B\kern-.05em{\sc i\kern-.025em b}\kern-.08em
    T\kern-.1667em\lower.7ex\hbox{E}\kern-.125emX}}
\begin{document}
\title{{\fontsize{24}{26}\selectfont{Communication\rule{29.9pc}{0.5pt}}}\break\fontsize{16}{18}\selectfont
Passive Decoupling of Two Closely Located Dipole Antennas}
\author{M.~S.~M.~Mollaei,~A.~Hurshkainen,~S.~Kurdjumov,~S.~Glybovski,~and~C.~Simovski
\thanks{M.~S.~M.~Mollaei, and C.~Simovski are with Department of Electronics and Nanoengineering, Aalto University, PO Box 15500, FI-00076, Finland (e-mail: masoud.2.sharifianmazraehmollaei@aalto.fi,konstantin.simovski@aalto.fi ). }
\thanks{A. Hurshkainen,~S.~Kurdjumov, and S. Glybovski are with the Department of Nanophotonics and Metamaterials, ITMO University, 197101, St. Petersburg, Russia (e-mail: a.hurshkainen@metalab.ifmo.ru, s.kurdjumov@metalab.ifmo.ru, s.glybovski@metalab.ifmo.ru)}}

\maketitle

\begin{abstract}
In this paper, we prove that two parallel dipole antennas can be decoupled by a similar but passive dipole located in the middle between them.
The decoupling is proved for whatever excitation of these antennas and for
ultimately small distances between them. Our theoretical model based on the method of induced electromotive forces is validated by numerical simulations and measurements. A good agreement between theory, simulation and measurement proves the veracity of our approach.
\end{abstract}


\section{Introduction}
\label{sec:introduction}
For many decades, usage of array antennas has the attention of researchers. Being employed in a variety of applications such as radars, Multi-Input Multi-Output (MIMO) systems and Magnetic Resonance Imaging (MRI) has made array antennas more interesting than any other time \cite{ref1,ref2}. Currently, decoupling of elements in the aforementioned applications has made the focus of research. For MIMO systems, a variety of techniques has been implemented and yet researchers strive to enhance those techniques \cite{ref3}--\cite{ref5}.
For antennas used in MRI, perhaps not ideal but sufficient, reliable and easily tunable decoupling is an important issue. In the transmission regime, decoupling of the array antennas prevents the parasitic cross-talks and  inter-channel scattering. In the reception regime, it prevents the noise correlation of channels that reduces the signal-to-noise ratio, one of key parameters of MRI. Finally, it makes the input impedances of two equivalent antennas 1 and 2 equivalent for arbitrary excitation magnitudes and phases. This simplifies the creation of the needed distribution of currents in the array elements and their impedance matching -- no need to engineer very expensive adaptive properties
in a decoupled array.

In the most of antenna array applications where the received signal is rather weak, the array elements cannot be decoupled in a straightforward way -- using screens or absorbing sheets. In last two decades, a very successful technique of the passive decoupling was developed for microwave antennas -- that based on the so-called \emph{electromagnetic band-gap} (EBG) structures. EBG structures are planar periodical structures -- microwave analogues of commonly known photonic crystals. Decoupling based on EBG structures works very well when the gap between two adjacent antennas may comprise a sufficient number of the EBG unit cells \cite{ref7}. This is, for example, the case of \cite{ref6}. In \cite{ref6} 3 unit cells of the EBG structure in the gap of the width $\lambda/10$ (where $\lambda$ is the operation wavelength) between two adjacent dipole antennas turned out to be sufficient for decoupling. This is, probably, the minimal gap for which the decoupling is possible using EBG structures.

Judging upon \cite{ref6,EBG,EBG1} the minimal amount of unit cells required for decoupling is equal 3. EBG structures with technically achievable miniaturization of the unit cell size allow one to place three unit cells
into a $\lambda/10$ gap \cite{EBG1}. However, no one knew technical solution of EBG structures allowing the unit cell miniaturization up to $\lambda$/100, and if the gap is as small as $d=\lambda$/30 one has to find other solutions.
The use of the adaptive circuitry based on operational amplifiers can be justified for radar systems but for MIMO and MRI applications it would be a very expensive and impractical way \cite{MRI}.
For the demands of such systems, the number of array elements needs to be increased, which necessarily leads to high inter-element coupling. Accordingly, one needs a passive decoupling for ultimately close antennas when $d\ll \lambda$/10  in terms of the operational wavelength.

%In \cite{ref8} one achieved this goal locating a scatterer called stacked magnetic resonator. It is a complicated system of tightly arranged complex-shape loops centered exactly in the middle between the dipoles. However, in the mentioned work no theoretical analysis of the decoupling was done; it was demonstrated as an experimental fact confirmed by extended numerical simulations. In fact, a specific situation was investigated in \cite{ref8}. In particular, the gap between antennas was as large as $\lambda$/12. We are interested in the much more challenging for the case when the distance between antennas is $\lambda$/30. In this case, a too small gap does not allow us to place the stacked magnetic resonator of \cite{ref8} in between antennas.

Elegant technical solutions were found for the case when the array elements are loop antennas. Since the mutual inductance of two coplanar loops is negative and that of two coaxial loops is positive, the loop array is performed of partially overlapping loops \cite{L,stas}. Another technical solution used for decoupling of loop antennas is capacitive decoupling \cite{C}. Unfortunately, for dipole antennas where the coupling is not purely inductive or capasitive, these methods are not applicable. In the present paper we aim for the passive decoupling, namely, in the arrays of dipole antennas. This problem becomes especially difficult when the gap width is as small as $d\ll \lambda$/10 and needs to be solved in two stages. We concentrate on two antennas -- the basic case of any type of array. On the next stage we will extend our theory to more antennas.

Our idea is to locate a passive scatterer in between two resonant dipoles. Basically, this idea is not fully new -- in work \cite{R} the authors revealed the decoupling offered to two resonant monopoles (vertical antennas fed by coaxial cables through a ground plane) by a similar monopole located in between them. The decoupling was obtained in presence of a human head phantom (stretched normally to the ground plane), the distance between the active monopoles was of the order of  $\lambda$/10.
Since the effects of the passive scatterer and of the body phantom were not studied separately, this technical solution was a heuristic finding. The same refers to the decoupling of two
loop antennas using a passive resonant loop in work \cite{R1,stas2}.

In Section II, we suggest a concept of the complete passive decoupling of two active antennas by adding a parasitic dipole. Complete decoupling means the suppression of the power flux
between the antennas obtained for arbitrary exciting magnitudes and phases. In Section III, a numerical investigation is presented and the results are compared with the theory.
In section IV we report an experiment that verifies both analytical and numerical results.

\section{Decoupling of Two Closely Located Dipoles}

\subsection{Reference Structure}

\begin{figure}
	\centering{\includegraphics[width=85mm]{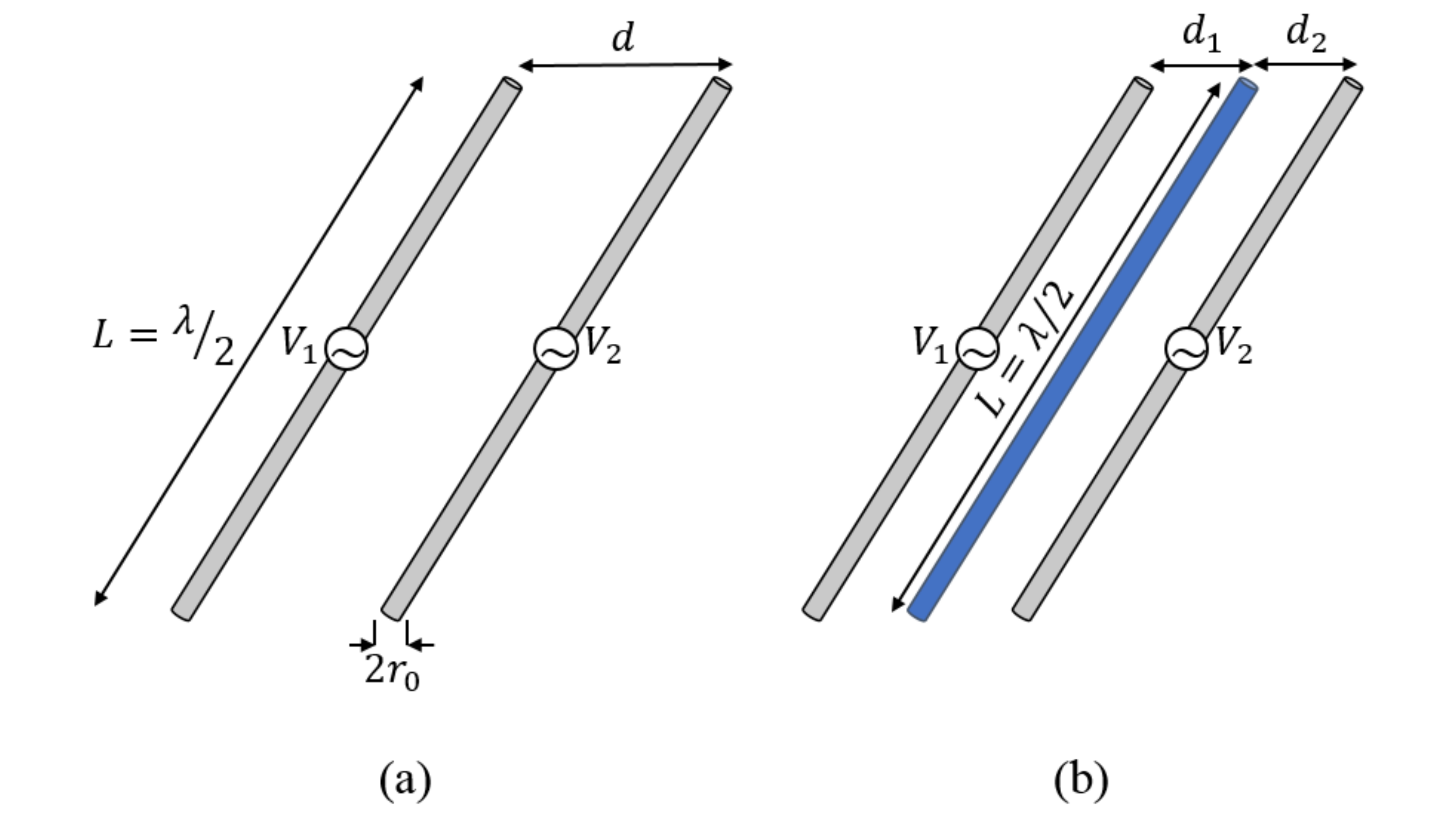}}
	\caption{(a) Two closely located active dipole antennas, (b) adding a passive dipole antenna between two active dipole antennas.}
	\label{fig1}
\end{figure}

Here we consider the interaction of two identical parallel dipole antennas separated by a gap $d$. Fig. \ref{fig1}(a) specifies the most interesting case when these dipoles are resonant.
Let them be fed by arbitrary voltages$\ V_1$ and$\ V_2$, respectively, and denote their self-impedances  as $\ Z_1{_1}= Z_2{_2}= Z$.

A system of Kirchhoff equations can be written in terms of the mutual impedances$\ Z_1{_2}$,$\ Z_2{_1}$ and, alternatively, via the shared (additional) impedances$\ Z^s_1$ and$\ Z^s_2$ of antennas:
\begin{equation}\label{cn1}
\begin{split}
{I_1}{Z_{11}} + {I_2}{Z_{12}} \equiv {I_1}\left( {{Z_{11}} + Z_1^s} \right) = {V_1}
\end{split}
\end{equation}
\begin{equation}\label{cn2}
\begin{split}
{I_1}{Z_{21}} + {I_2}{Z_{22}} \equiv {I_2}\left( {{Z_{22}} + Z_2^s} \right) = {V_2}
\end{split}
\end{equation}
Here currents $I_{1,2}$ refer to the centers (feeding point) of the dipoles.
Mutual impedances, as follows from the reciprocity, are the same$\ Z_{12}= Z_{21}\equiv Z_M$. The shared impedances of two antennas$\ Z^s_1$ and$\ Z^s_2$ are different if$\ I_1\neq I_2$. They express the electromotive forces induced in antenna 1 by antenna 2 (and vice versa) normalized to the current in antenna 1 (or 2). Let us write $\ I_2=\alpha I_1$, where the coefficient
$\alpha$ is an unknown complex value. If it is different from unity$\ Z^s_1=\alpha Z_M$ and$\ Z^s_2=Z_M/\alpha$ are different. Respectively, input impedances of the equivalent antennas$\ Z_1 = Z + Z^s_{1}$ and$\ Z_2 = Z + Z^s_{2}$ are also different. Therefore, mutual coupling of two antennas means that the relation $\alpha\ne 1$ between the currents is different
from the relation between their voltages $\ V_2 \ne \alpha V_1$.

For decoupled antennas we would have$\ V_2/V_1 = I_2/I_1$ for whatever $\alpha$. Expanding this result to an array with $N>2$ elements we observe that the distribution of currents in its elements repeats that of applied voltages. All we need for it is to nullify the electromotive force induced in the given antenna by other active antennas. It is possible if the impact of our passive scatterers compensates the impedance shared by the given antenna with the other active antennas of the array.

\subsection{Structure with a Passive Element}

In this part, we analytically prove that compensation of the electromotive force induced by antenna 1 in antenna 2 and vice versa
is possible at a certain frequency if dipole scatterer 3 is introduced as it is shown in Fig. \ref{fig1}(b). In the regime of decoupling the input impedances of both antennas 1 and 2 are equivalent, and $\ V_2 = \alpha V_1$ if $\ I_2 = \alpha I_1$.
%This implies that the impedance shared by antenna 1 with antenna 2 (and vice versa) is compensated by a part of the impedance shared by antenna 1 with scatterer 3.

The impedance of antenna 1 shared with both antenna 2 and scatterer 3 is as follows:
\begin{equation}\label{cn3}
\begin{split}
Z_1^s = {Z_M}\frac{{{I_2}}}{{{I_1}}} + {Z_{13}}\frac{{{I_3}}}{{{I_1}}}
\end{split}
\end{equation}
In the decoupling regime it must be equal to the impedance of the second antenna shared with antenna 1 and scatterer 3:
\begin{equation}\label{cn4}
\begin{split}
Z_2^s = {Z_M}\frac{{{I_1}}}{{{I_2}}} + {Z_{23}}\frac{{{I_3}}}{{{I_2}}}
\end{split}
\end{equation}
Here$\ Z_{13}$ and$\ Z_{23}$ are mutual impedances of, respectively, antennas 1 and 2 with scatterer 3 and$\ Z_M \equiv Z_{12}$.

Current$\ I_3$ in the center of scatterer 3 is induced by primary currents$\ I_1$ and$\ I_2$ (where by definition $I_2=\alpha I_1$). Due to linearity of electromagnetic interaction, the current induced in a scatterer by a primary current must be proportional to this primary current, and we have
\begin{equation}\label{cn5}
\begin{split}
{I_3} = {\xi _{13}}{I_1} + {\xi _{23}}{I_2}
\end{split}
\end{equation}
Here$\ \xi _{13}$ and$\ \xi _{23}$ are certain coefficients, determined by the system geometry and independent on currents$\ I_1$ and$\ I_2$. With these coeffcients the equivalence of \eqref{cn3} and \eqref{cn4} can be written in a form
\begin{equation}\label{cn6}
\begin{split}
{Z_M}\alpha  + {Z_{13}}\frac{{{\xi _{13}}{I_1} + {\xi _{23}}{I_2}}}{{{I_1}}} = {Z_M}\alpha  + {Z_{23}}\frac{{{\xi _{13}}{I_1} + {\xi _{23}}{I_2}}}{{{I_2}}} =\\ \left( {{Z_M} + {\xi _{23}}{Z_{13}}} \right)\alpha  + {\xi _{13}}{Z_{13}} = \left( {{Z_M} + {\xi _{13}}{Z_{23}}} \right)\frac{1}{\alpha } + {\xi _{23}}{Z_{23}}
\end{split}
\end{equation}
If the distances$\ d_1$ and$\ d_2$ are equivalent (the scatterer is symmetrically located)$\ Z_{13} = Z_{23}$ and$\ \xi _{13} = \xi _{23}$. In this case, \eqref{cn6} becomes an identity when$\ \alpha = 1$. Of course, this equivalence of the input impedances does not mean their decoupling. For arbitrary $\alpha$ \eqref{cn6} is satisfied when
\begin{equation}\label{cn7}
\begin{split}
{Z_M} + {\xi _{23}}{Z_{13}} = {Z_M} + {\xi _{13}}{Z_{23}} = 0.
\end{split}
\end{equation}
%
%that is easier to implement when$\ Z_{13} = Z_{23}$ and$\ \xi _{13} = \xi _{23}$.
Equation \eqref{cn7} is the needed condition of complete decoupling. If it is satisfied input impedance of antenna 1,$\ Z_1 = Z + Z^s_1$ does not depend on$\ I_2$ and vice versa,
the shared impedances of antennas 1 and 2 do not depend on$\ I_1$ and$\ I_2$ and are equal to each other:	
\begin{equation}\label{cn8}
\begin{split}
{Z^s_1} = {Z^s_2} = {\xi _{13}}{Z_{13}} = {\xi _{23}}{Z_{23}}
\end{split}
\end{equation}
Equations \eqref{cn7} and \eqref{cn8} mean that the electromotive force induced in antenna 1 by antenna 2 (and vice versa) is always compensated by a part of the electromotive force
induced in them by scatterer 3. If there is no mutual coupling of 1 and 2 via the electromotive force, there is no power flux between them.

To express$\ \xi _{13}$ through$\ Z_M \equiv Z_{12} = Z_{21}$, denote the self-impedance of scatterer 3 as $\ Z_0$.
For calculating$\ \xi _{13}$ and$\ \xi _{23}$, we need two different scenarios of excitation. In the first scenario, we assume that dipole 1 is active ($V_1 = V$)
while dipole 2 is as passive as dipole 3 ($V_2 = 0$). We may write the system of Kirchhoff's equations for our three dipoles as follows:
\begin{subequations}
	\begin{equation}
	{I_1}Z + {I_2}{Z_M} + {I_3}{Z_{13}} = V,  \label{cn9a}
	\end{equation}
	\begin{equation}
	{I_1}{Z_M} + {I_2}Z + {I_3}{Z_{23}} = 0,  \label{cn9b}
	\end{equation}
	\begin{equation}
	{I_1}{Z_{13}} + {I_2}{Z_{23}} + {I_3}{Z_0} = 0,  \label{cn9c}
	\end{equation}
\end{subequations}
%
%Solving the above system, a set of currents named$\ {I^{'}}_1$,$\ {I^{'}}_2$ and$\ {I^{'}}_3$ is achieved for the antennas.

%Similarly, by applying the system of Kirchhoff equations to the antennas and solving it, another set of currents named$\ {I^{''}}_1$,$\ {I^{''}}_2$ and$\ {I^{''}}_3$ is achieved. Substituting these sets of currents into \eqref{cn5}, we have:
%
%\begin{subequations}
%	\begin{equation}
%	{I^{'}_3} = {\xi _{13}}{I^{'}_1} + {\xi _{23}}{I^{'}_2}  \label{cn10a}
%	\end{equation}
%	\begin{equation}
%	{I^{''}_3} = {\xi _{13}}{I^{''}_1} + {\xi _{23}}{I^{''}_2}  \label{cn10b}
%	\end{equation}
%\end{subequations}
\noindent
and easily obtain:
\e
\kappa_{13}\equiv{I_3\over I_1} ={Z_{M}Z_{23}-ZZ_{13}\over ZZ_0-Z_{23}^2}\l{kap1}\f
In the second scenario, we assume that dipole 2 is active and two other dipoles are passive:$\ V_1 = 0$,$\ V_2 = V$.
Then we obtain
\e
\kappa_{23}\equiv{I_3\over I_2} ={Z_{M}Z_{13}-ZZ_{23}\over Z_0Z-Z_{13}^2}\l{kap2}\f
With substitutions of \r{kap1} and \r{kap2}, condition \eqref{cn7} can be rewritten in terms of mutual and self-impedances:
\begin{equation}\label{cn11}
\begin{split}
{Z_{13}} = \sqrt {{Z_0}{Z_M}} = {Z_{23}}.
\end{split}
\end{equation}
This condition is that of complete decoupling of two active antennas by a passive one. Our term \emph{complete decoupling} means the following:
if antennas 1 and 2 are fed by arbitrary voltage sources $V_1$ and $V_2$ with internal impedances $Z_{i1}$ and $Z_{i2}$ we still have no mutual coupling.
The impedances $Z_1$ and $Z_2$ connected to the voltages $V_1$ and $V_2$ comprise the antenna self-impedance $Z$, the source impedances $Z_{i1,i2}$ and
the shared impedances $Z_{1,2}^s$ of antennas 1 and 2. If our condition \eqref{cn11} is satisfied we obtain from Kirchhoff's equations for the structure depicted in
Fig.~\ref{fig1}(b) the equivalence of the shared impedances $Z_1^s=Z_2^s$. To obtain it we substitute in \eqref{cn9a} $Z\rightarrow Z+Z_{i1}$ and $V \rightarrow V_1$, and
in Eq. \eqref{cn9b} $Z\rightarrow Z+Z_{i2}$ and $0 \rightarrow V_2$. Then we substitute for $Z_{13}$ and $Z_{23}$ relation \eqref{cn11},
solve Kirchhoff's equations for $I_1$ and $I_2$ and obtain $V_1/I_1-Z_{i1}=V_2/I_2-Z_{i2}$ i.e. $Z_1-Z_{i1}=Z_2-Z_{i2}$. Since
$Z_1=Z+Z_{i1}+Z_1^s$ and $Z_2=Z+Z_{i2}+Z_2^s$ it means $Z_1^s=Z_2^s$. This equivalence of the shared impedances in the case when two arbitrary different
generators are connected to our antennas may result only from their decoupling.

Now, we have to prove the feasibility of our condition \eqref{cn11}. In our derivations
we did not assume the exact symmetry of the location of scatterer 3. However, it is required by Eq. \eqref{cn11}:$\ {Z_{13}} = {Z_{23}}$.
Let us show that \eqref{cn11} is feasible when scatterer 3 is the same resonant dipole located between dipoles 1 and 2 in the same plane.
This case is shown in Fig. \ref{fig1}(b).

\subsection{Decoupling of Two Half-Lambda Dipoles}	

If $\ Z_0 = Z$ (self-impedances of all our dipoles are equivalent) the decoupling condition yields to:
\begin{equation}
	{Z_{13}}^2 = {Z}{Z_M}  \label{cn12}
\end{equation}

Formulas for the self-impedance of a dipole antenna located in free space and for the mutual impedances of two parallel dipole antennas are well known. Standard formulas represent converging power series \cite{ref9}, closed-form combinations of integral sine and cosine functions \cite{ref10}, or polylogarithm functions \cite{ref11}. However, for our purposes we do not need these long formulas. We consider the special case of resonant dipoles and may use simple approximate relations, valid in the vicinity of the antenna resonance. First, let us recall that the straight-wire resonance holds at a frequency that is slightly lower than the one at which$\ L = \lambda / 2$. By definition, at the resonance frequency$\ \omega_0$ the reactive part of the self-impedance vanishes. For a perfectly conducting wire of length $L$ and cross section radius$\ 10^{-5} \lambda < r_0 < 10^{-3} \lambda$ (this interval of values for $r_0$ is assumed below) the resonant wave number$\ k_0 \equiv 2 \pi / \lambda$ equals to$\ 0.992 \pi / L$ \cite{ref10,ref11,ref12}. At this frequency, the input resistance of the dipole is equal$\ R_0 \approx 70$ Ohm \cite{ref10,ref11,ref12}.

A very simple relation for the mutual impedance of two parallel antennas performed of wires with radius$\, r_0 < 10^{-3} \lambda$ separated by a gap$\ d<L$ was heuristically obtained in \cite{ref13}. In our notations this relation can be written for both ${Z_{12}} = {Z_M}$ and ${Z_{13}} = {Z_{23}}$ as follows:
\begin{subequations}
	\begin{equation}
		{Z_M} = \frac{\eta }{{24\pi kd}}{e^{ - jkd}}  \label{cn13a}
	\end{equation}
	\begin{equation}
		{Z_{13}} = \frac{\eta }{{12\pi kd}}{e^{ - jkd / 2}}  \label{cn13b}
	\end{equation}
\end{subequations}
Here it is taken into account that the distance between 1 and 3 is $d/2$ and it is denoted$\ \eta \equiv \sqrt {({\mu}_0 / {\epsilon}_0)}$ (free-space impedance). Comparison with the known data for the mutual impedance of two parallel identical dipoles \cite{ref11}, shows that Eqs. \eqref{cn13a} and \eqref{cn13b} are sufficiently accurate when $r_0\ll d \ll \lambda$ at frequencies located within the half-lambda resonance band. This resonance band
is defined via the relative detuning $\gamma$ as follows:$\ |\gamma|  \equiv  |(\omega  - {\omega _0}) / {\omega _0}|\le 0.01$.
If we show that Eq. \eqref{cn12} holds at a frequency within this band the use of approximations  \eqref{cn13a} and \eqref{cn13b} for the mutual impedances will be justified.

The input reactance $X$ of a dipole within the above-defined resonance band is negligibly small compared to the input resistance $R$ and we have $Z\approx R$ in \eqref{cn12}. The dependence of $R$ on the relative detuning in the resonance band of a half-lambda dipole can be modelled by a linear function \cite{ref14}:
\begin{equation}
	Z = {R_0}\left( {1 + \beta \gamma } \right)  \label{cn14}
\end{equation}
where$\ \beta \approx 59$. Substituting Eqs. \eqref{cn13a} -- \eqref{cn14} into \eqref{cn12}, we obtain:
\begin{equation}
	\begin{split}
		{R_0}\left( {1 + \beta \gamma } \right)\frac{\eta }{{24\pi kd}}{e^{ - jkd}} =  \frac{{{\eta ^2}{e^{ - jkd}}}}{{{{\left( {12\pi kd} \right)}^2}}}  \label{cn15}
	\end{split}
\end{equation}
In this equation, complex exponentials cancel out. Substituting$\ \eta = 120$ Ohm, we simplify \eqref{cn15} to a form:
\begin{equation}
	70\left( {1 + 59\frac{{k - {k_0}}}{{{k_0}}}} \right) = \frac{{20}}{{kd}}  \label{cn16}
\end{equation}
This is clearly a feasible condition. For example, for two dipoles of length $L=50$ cm performed of a wire with radius $r_0=1$ mm (such a dipole resonates, in accordance to the analytical theory, at 298 MHz) separated by a gap $\ d = 3$ cm, that is centered by a wire of the same radius and length, Eq. \eqref{cn16} is satisfied when$\ \gamma  = (\omega  - {\omega _0}) / {\omega _0} \approx 0.007$, i.e. at frequency$\ \omega  = 1.007{\omega _0}\approx 300$ MHz. Thus, the prerequisite of \eqref{cn13a} and \eqref{cn13b} is respected: decoupling holds within the resonance band.

\section{Numerical Verification}
\label{sec:guidelines}

 In this part, the proposed method is validated numerically. The simulation has been carried out using CST Microwave Studio, Time Domain solver.
 %The most important reason of using this solver is its calculation speed, which is much faster than other solvers for the situation when size of antenna is in the order of wavelength.
%	
\begin{figure}
	\centering{\includegraphics[width=80mm]{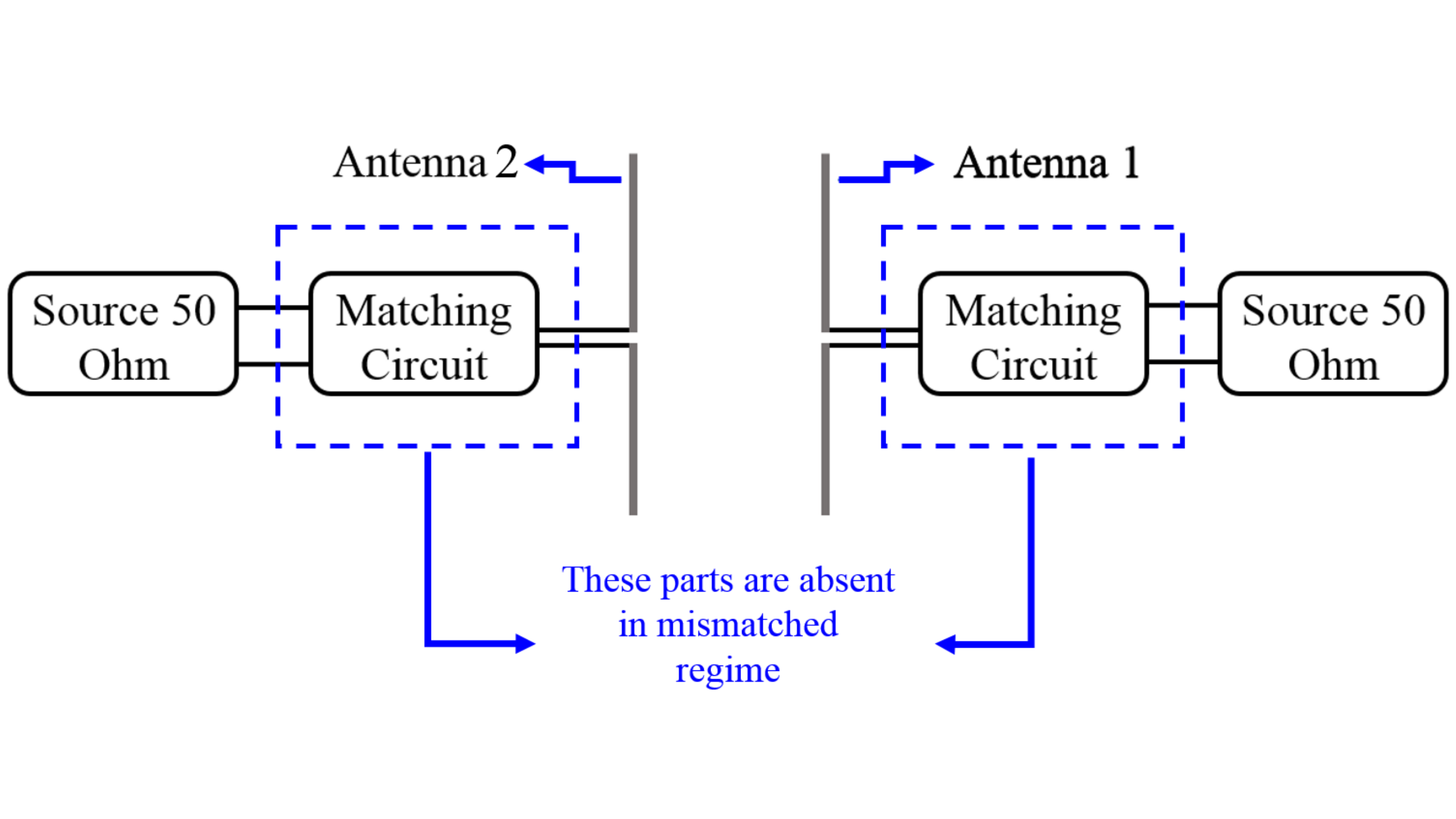}}
	\caption{Reference structure of two dipole antennas with ideal lossless matching circuits (removed in the mismatched regime).}
	\label{fig2}
\end{figure}

As a reference, we consider the system of two dipole antennas performed of copper wires with geometric parameters$\ L=$ 500 mm, $\ r_0=$ 1 mm separated by the distance $d=$ 3 cm. We simulate S-parameters of the system assuming our dipoles to be performed of copper wires. In these simulations, dipoles 1 and 2 were excited by lumped ports either through matching circuits or without them.
This schematic of the reference antenna system is shown in Fig.~\ref{fig2}. In the matched case lumped voltage sources $V_{1,2}$ with internal resistances $R_{i1,i2}=$ 50 Ohms are connected to the dipoles centers through a lossless LC circuit whose parameters are chosen so that its reactance $X_{i1,i2}$ compensates the antenna input reactance at 300 MHz and the input resistance transforms into 50 Ohms. In this case in absence of scatterer 3 the coupling is maximal ($|S_{12}|=-$1.2 dB) at 293 MHz that is an evident consequence of the maximal antenna currents at this frequency. The band of matching in the reference structure defined by condition $|S_{11}|\le -$20 dB  is equal [292.7,293.3] MHz (relative bandwidth 0.2\%). It is desirable to keep this band in the regime of decoupling. Indeed, a scatterer located so closely to our dipoles obviously brings an extra mismatch. This mismatch can be compensated by adjusting the matching circuit. However,ideal matching with a reactive circuit is possible at a single frequency. Therefore, the decoupling element may severely shrink the band of the antenna matching that may become narrower than that of the transmitted signal. Then one has to introduce losses in the matching circuit that means the decrease of the antenna efficiency compared to the reference structure. Moreover, the band of decoupling can be even narrower than the band of lossless matching. Our Eq. \eqref{cn16} allows us to find a single frequency of decoupling and tells nothing about its matching band. These fine issues can be hardly covered by our approximate theory and we clarify them in extensive numerical simulations.

\begin{figure}
	\centering{\includegraphics[width=100mm]{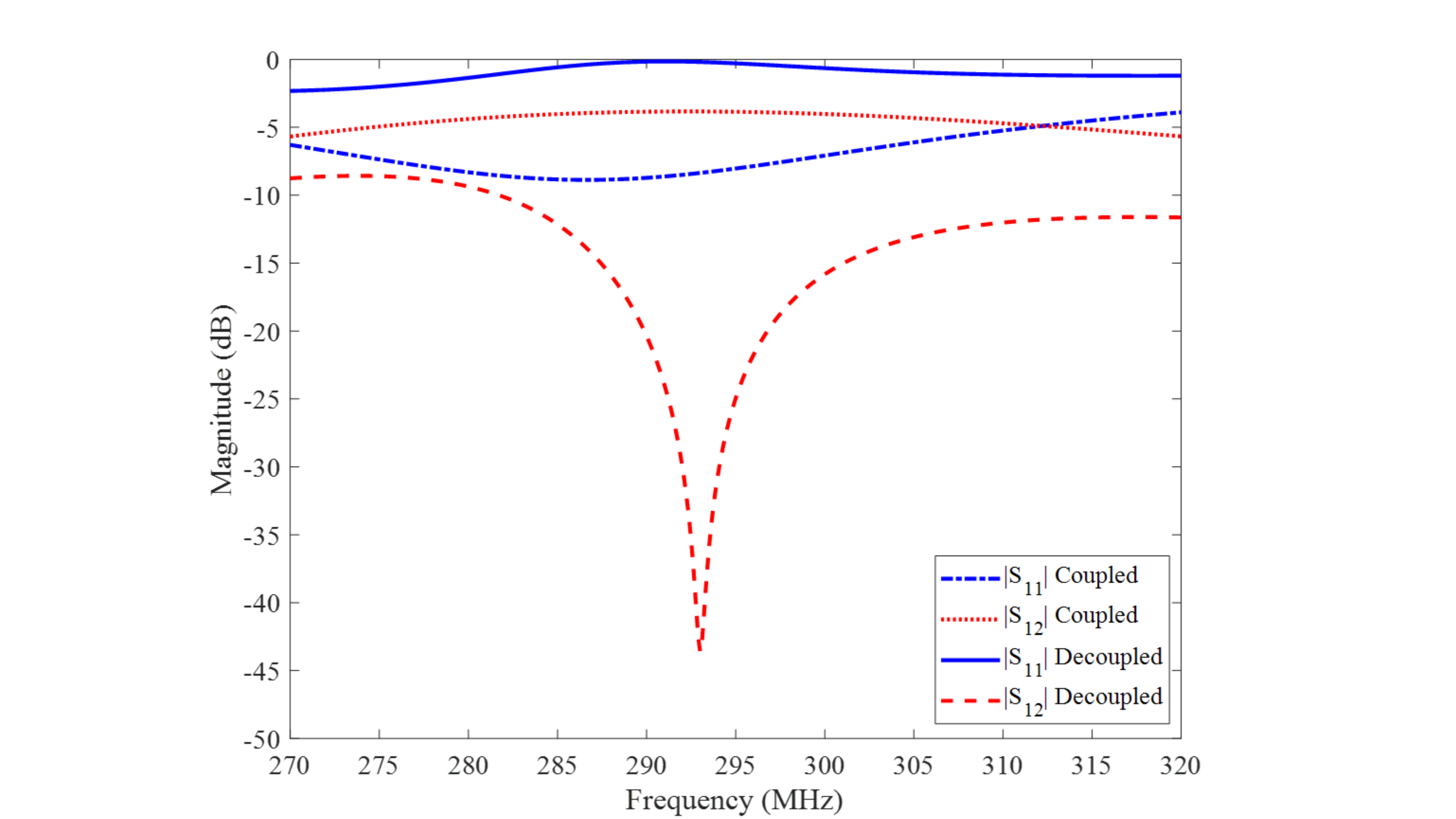}}
	\caption{Simulation results for the reference and decoupled structures. Matching circuits are absent.}
	\label{fig4}
\end{figure}

\begin{figure}
	\centering{\includegraphics[width=100mm]{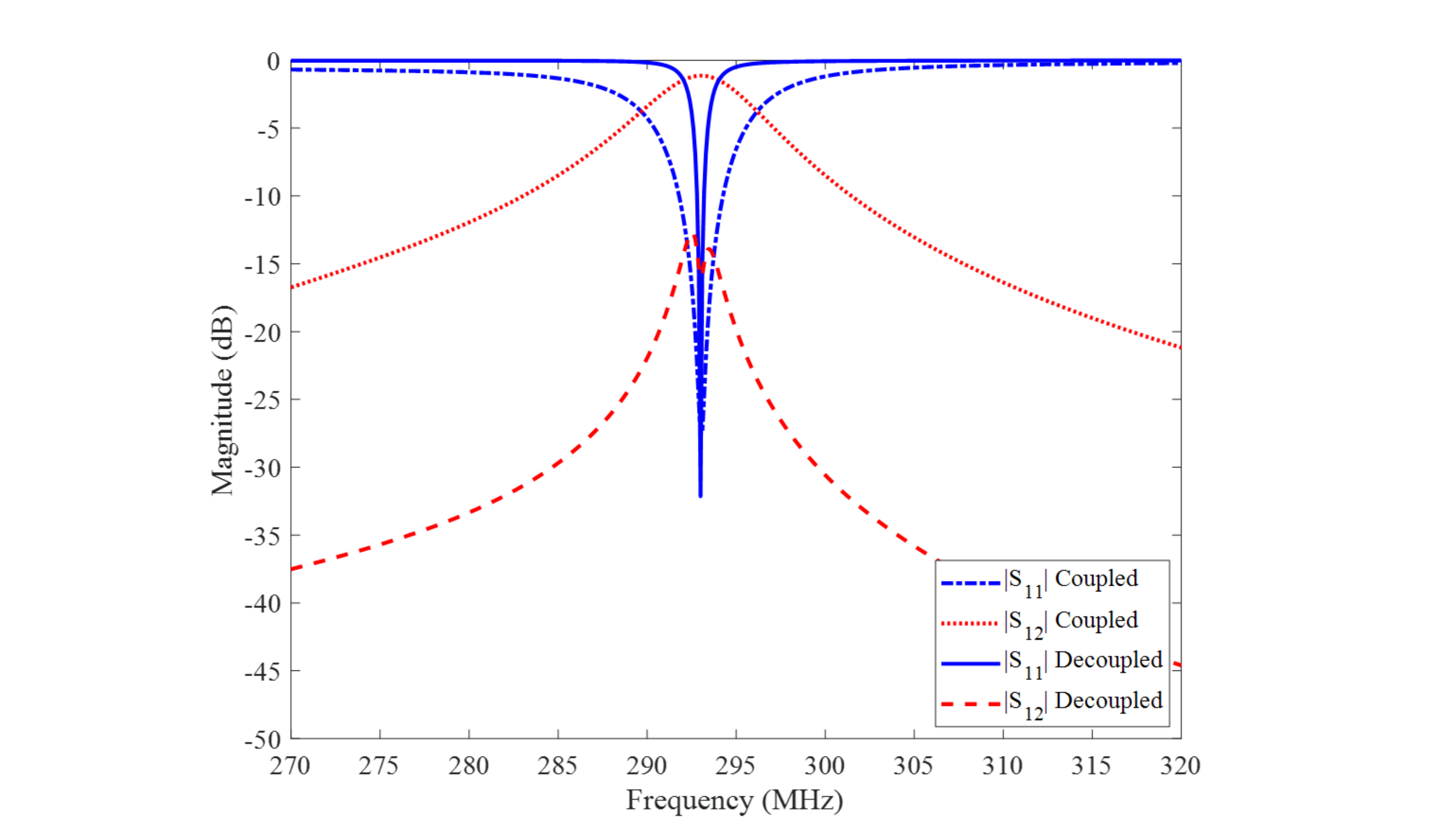}}
	\caption{Simulated S-parameters of the reference and decoupled structures. Matching circuits are present.}
	\label{fig3}
\end{figure}

In Figs.~\ref{fig4} and~\ref{fig3} we depict  the results of  numerically calculated S-parameters of two dipoles in the presence of a passive scatterer without and with a matching circuit accordingly.
Our simulations confirmed the prediction of the analytical model that decoupling of our dipoles 1 and 2 is granted by the straight wire (scatterer 3) of the same length and radius, located in the middle between 1 and 2. The decoupling is complete because it is achieved in both matched and mismatched cases. In the matched case $X_{i1,i2}$ is the value of the order of $R_{i1,i2}$, whereas
in the mismatched case $X_{i1,i2}=0$. Therefore, the ratio of currents in the passive and active antennas ($\alpha=I_2/I_1$) is different in these two cases.
However, in both matched and mismatched cases we have obtained a local minimum of $S_{12}$ at the same frequency. This is an evidence of the complete decoupling.
Of course, in both these cases $S_{12}$ cannot exactly vanish. Though our analytical model is approximate, and the simulation shows that the reachable isolation is incomplete, it can be seen that the complete decoupling condition is still valid. We note that in the most of applications the isolation $|S_{12}|<-20$ dB of two matched antennas is sufficient.

As expected, the decoupled structure in the mismatched case manifests an extra mismatch compared to the reference one.
In the reference structure the mutual coupling is not so harmful for matching. Mutual impedance of two closely located
($d\approx \lambda/30$) dipoles in their resonance band has absolute values within the limits [20,50] Ohms (see e.g. in \cite{ref12}), and the shared impedance $Z_1^s$ is smaller 
than the mutual impedance because dipole 2 is passive and, therefore, $|\alpha|<1$. Consequently, for the reference structure we see a broadband (though poor)
matching with the resonance frequency 293 MHz, almost equal to that of an individual dipole. In the decoupling structure dipole 3 is distanced
by $d/2 = \lambda/60 = 1.5$ cm from our active dipole 1. For a so small distance, the mutual resistance of two half-wave dipoles approaches the self-resistance
and $\kappa_{13}$ is almost real and negative. Therefore, dipole 3 is excited in the opposite phase and the shared resistance of dipole 1 almost 
compensates its self-resistance. As a result, at the decoupling frequency 293 MHz the absolute value of the input impedance of dipole 1 turns out to be much smaller than $R_{i1}=$50 Ohm
which implies a very strong mismatch we observe for the decoupled structure in Fig.~\ref{fig4}.

However, whatever nonzero input impedance can be always matched at a single frequency using a lossless matching circuit. 
In Fig.~\ref{fig3} we show the S-parameters of the matched system calculated in absence and in presence of the scatterer. 
Again the decoupling holds at 293 MHz because within the band 292.7-293.3 MHz the minimum of $S_{12}$ (equal to -20 dB) is achieved at 293 MHz.
Beyond this band $S_{12}$ also decreases versus detuning, but this is a consequence of the mismatch, and is not decoupling.

From Fig.~\ref{fig4} and~\ref{fig3} we conclude the minimum of $S_{12}$ at 293 MHz does not depend on the currents in antennas 1 and 2 and represents an evidence of the complete decoupling.
Notice that the result for the decoupling frequency 293 MHz fits very well the prediction of the analytical model.
The only drawback of our decoupling is squeezed operational band because our dipole 3 brings a strong mismatch.
Namely, in accordance to Fig.~\ref{fig3}, the relative band of matching defined via $|S_{11}|\le -20$ dB shrinks from 0.2\% (reference structure) to 0.05\%.
The band of matching is practically equal to the band of decoupling defined on the level $|S_{12}|=-15$ dB.

\section{Experimental Validation and Discussion}

\begin{figure}
\centering{\includegraphics[width=90mm]{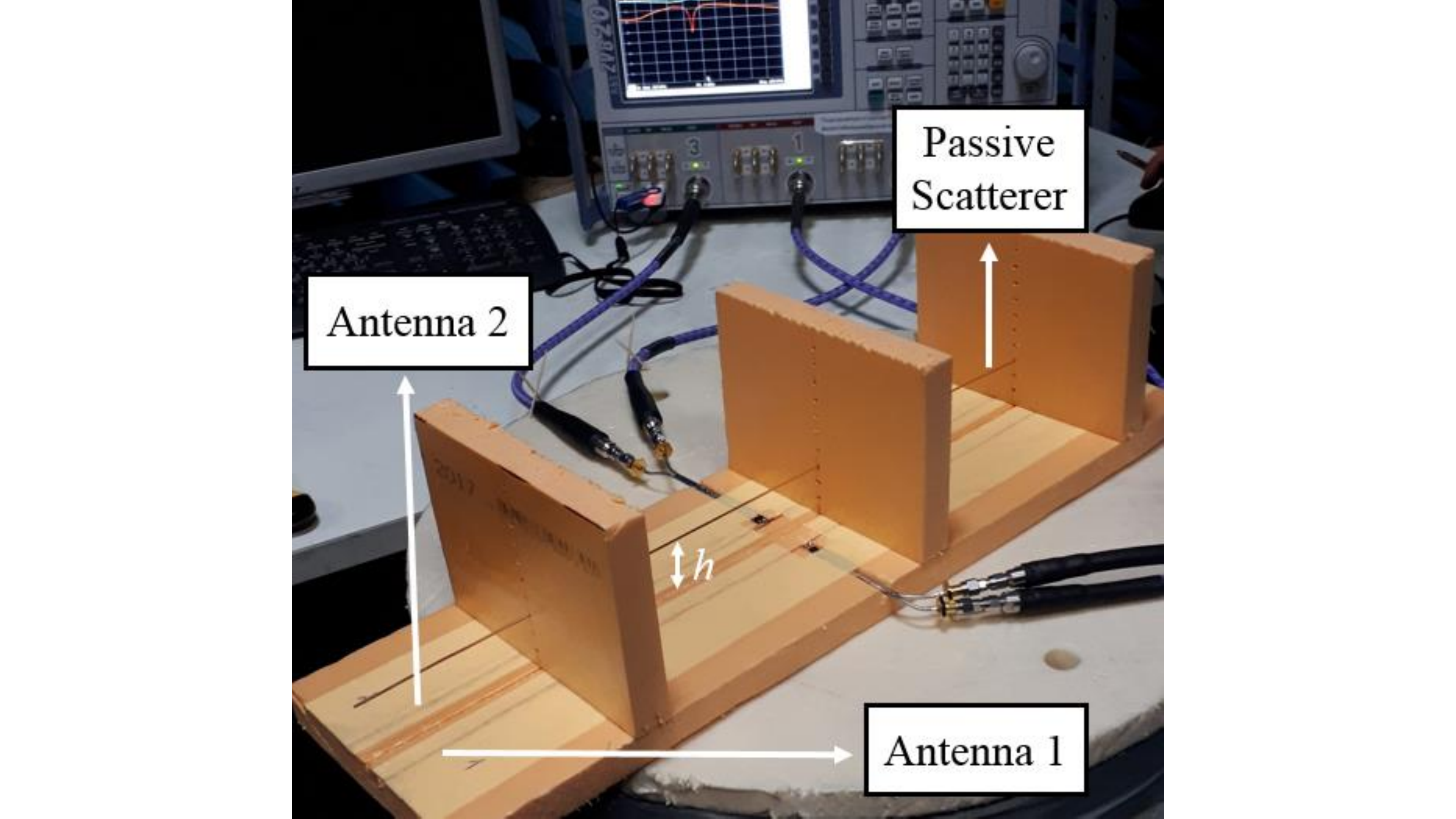}}
%\centering\subfigure[]{\includegraphics[width=70mm]{setup.png}\label{fig51}}
%\subfigure[]{\includegraphics[width=70mm]{new1.png}\label{fig52}}
 %\centering{\includegraphics[width=75mm]{setup1.png}}
 \caption{%(a) --
 Measurement setup comprises two dipole antennas connected to a VNA and a dipole scatterer. The antenna system
 is supported by a foam board wrapped with the paper. The board design allowed to raise the passive scatterer.
 %(b) -- Foam plates wrapped with the paper are used in order to change the scatterer height $h$.
 }
 \label{fig5}
\end{figure}

For the experimental validation of our theory we built a setup whose general views are pictured in Fig. \ref{fig5}. It comprised active dipoles 1 and 2 performed of a copper wire
with $L=$500 mm and $r_0=1$ mm split at the center with a small (0.6 mm) antenna gap. A vector network analyzer (VNA) Rohde Schwarz ZVB20 was connected to the arms of the dipoles
through a logical symmetric ports with the wave impedance 100 Ohm each. In this scheme two identical coaxial cables are used to connect each of the dipoles. The symmetric
connection allowed us to measure the S-parameters properly removing the cable effect in the whole frequency range. The distance between the dipole antennas was $d=3$ cm. In the decoupling regime the similar copper wire ($L=500$ mm, $r_0=1$ mm but without a central split) was located in the middle of the antenna structure.
%The change of the length of antennas with respect to previous simulations is not important. In any case it is impossible to exactly reproduce the previous results experimentally because
%the actions of the plastic board and of the finite antenna gap are not negligible (though small). Therefore, we repeated the simulations for the case $L=$505 mm taking in account the
%plastic support.

\begin{figure}
	\centering{\includegraphics[width=100mm]{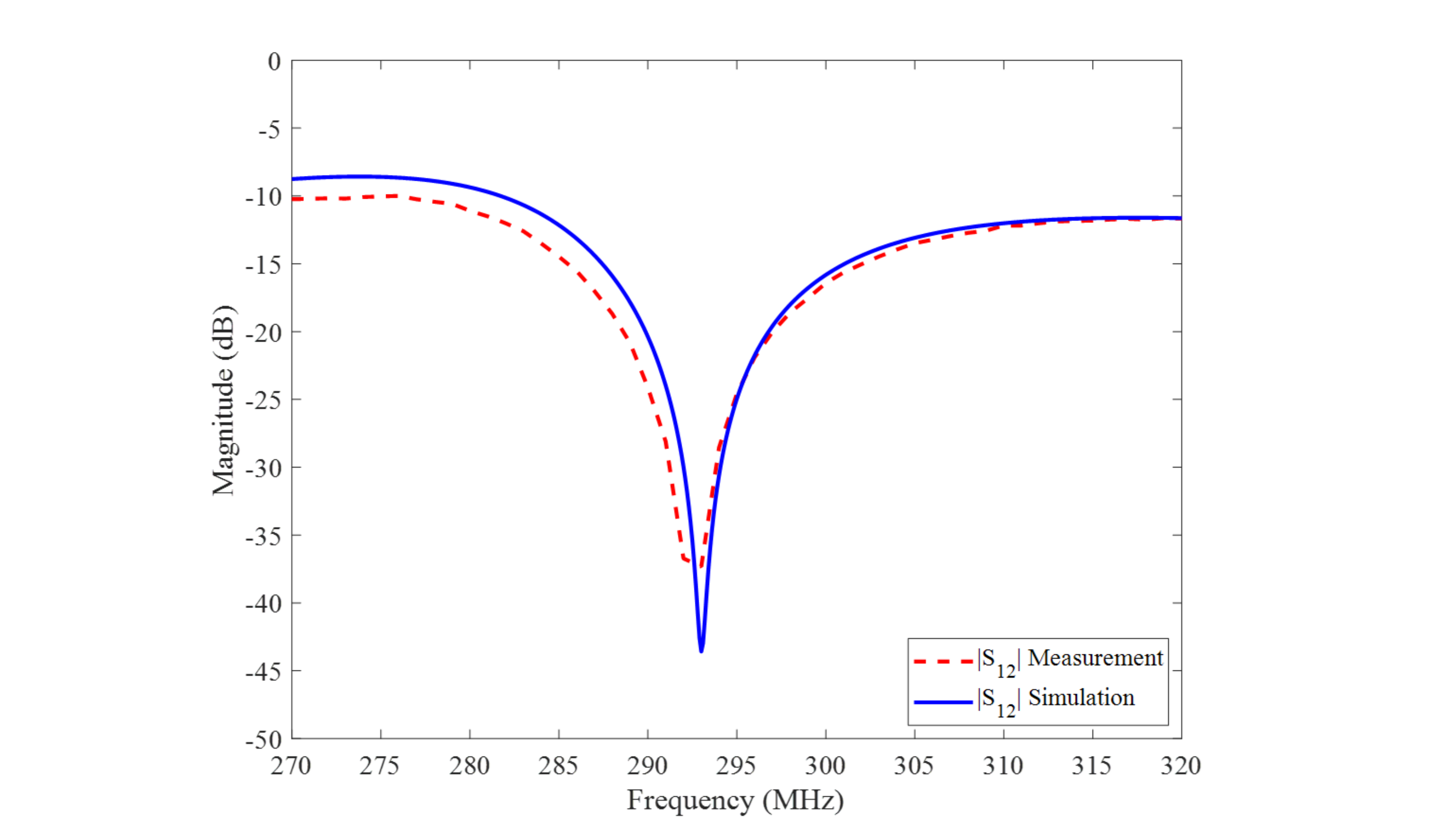}}
	\caption{Comparison of simulation and measurement results for the decoupled structure in mismatched regime.}
	\label{fig6}
\end{figure}

\begin{figure}
	\centering{\includegraphics[width=100mm]{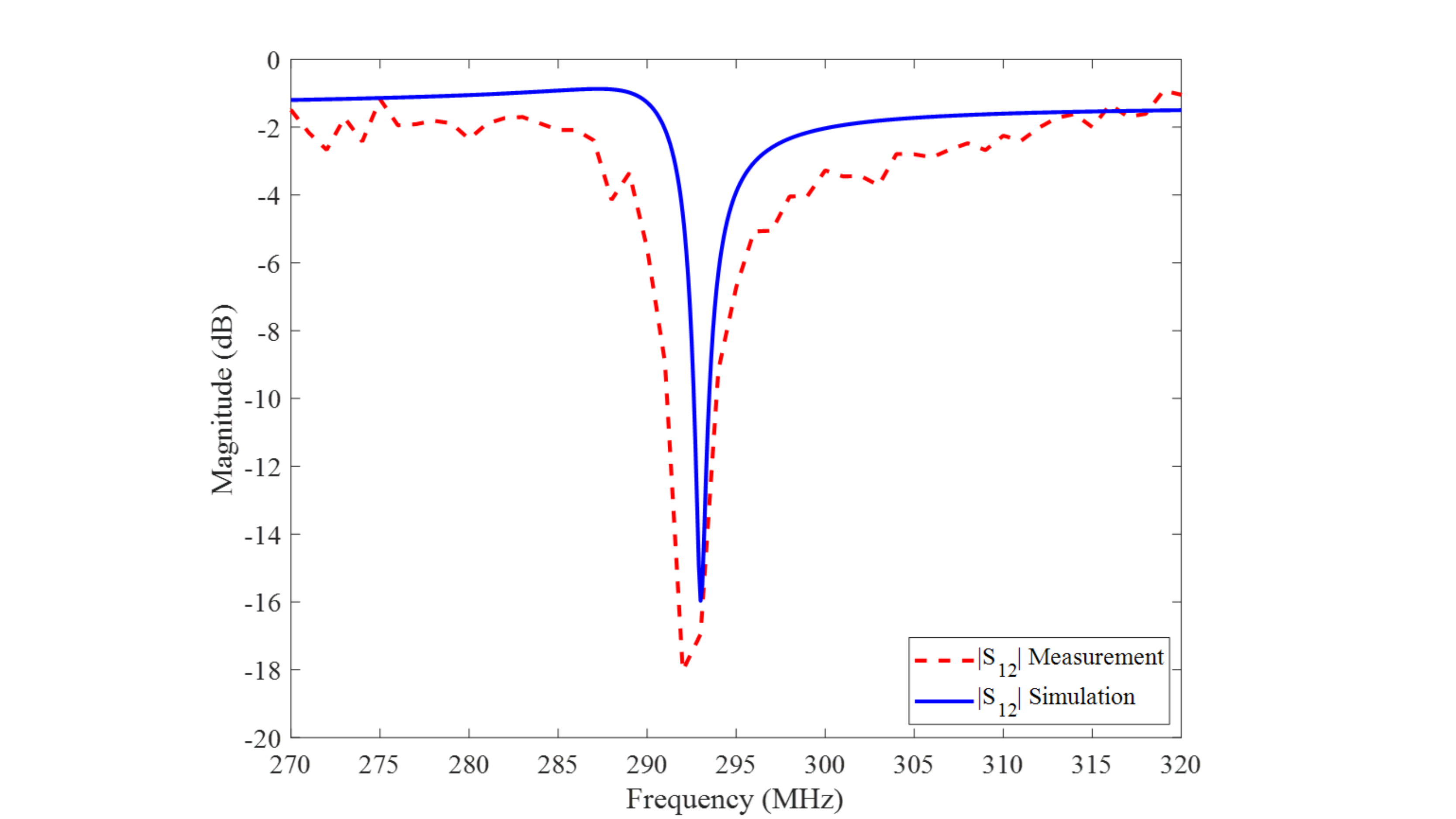}}
	\caption{Comparison of simulation and measurement results for the decoupled structure when an ideal dual side matching of the two port is performed.}
	\label{fig7}
\end{figure}

A mechanical support was a board of foam wrapped with paper that allowed us to raise scatterer 3 to a height $0< h \le 5$ cm over the plane of dipoles 1 and 2. Varying $h$ in both measurements and simulations, we found complete decoupling is possible also for $h\ne 0$. Measurements of S-parameters for the reference case (without scatterer 3) have shown an excellent agreement with these simulations.
For the structure with the decoupling scatterer the agreement is still acceptable as one can see in Figs. \ref{fig6} and \ref{fig7}. These plots correspond to $h=0$.

%In both measurements and simulation the band of matching is nearly equal to the band of decoupling.
%In the measurements this band was wider that is, definitely, related to losses that were not taken into account in our simulations.

In the measurement whose result is depicted in Fig.~\ref{fig6} there is no matching circuits connected to antennas 1 and 2. In our simulations
of this mismatched case the minimum of $S_{12}$ occurs at 293 MHz, which is exactly the same as experiment.
The same frequencies of the minima of $S_{12}$ (simulated and measured ones) keep for the matched case. The corresponding plot is shown
in Fig.~\ref{fig7}, where the ideal two-port matching is supposed at all frequencies. The coincidence of the frequencies of these minima in Figs.~\ref{fig6} and \ref{fig7} is the evidence of the complete decoupling.

Here, instead of real matching circuits we used MATLAB code based on the method from \cite{mat} which allows us to normalize $S_{12}$ (was measured in the mismatched regime
and depicted in Fig.~\ref{fig6}) to the accepted power considered as a symmetric passive two-port (dual-side matching). Mathematically, the obtained result is equivalent to the presence of an ideal matching circuit transforming the
antenna input impedance into 50 Ohms at each of the plotted frequencies. This approach was dictated by the necessity to measure $S_{12}$
for different $h$ allowing us to avoid the fabrication of a tunable matching circuit and gives the reachable isolation between the dipoles at all frequencies, for which each of those a lossless matching circuit could be individually constructed.

Simulations and measurements of $S_{12}$ for $h$ varying in the limits $0\le h\le 50$ mm have shown no complete decoupling for $h\ne 0$.
Though the deepest minimum of $S_{12}=-$17 dB in the matched case was obtained for $h=10$ mm, this was not our complete decoupling, because
this minimum corresponds to frequency 294.2 MHz, whereas in the mismatched case for $h=10$ mm $S_{12}$ attains the minimum at 293.2 MHz.
Only if $h=0$ the frequency of the minimum of $S_{12}$ keeps the same in both matched and mismatched cases which goes along with the theory.

\section{Conclusion}
In this work we have comprehensively studied the passive decoupling of two active dipole antennas by a single passive dipole scatterer. We aimed the complete decoupling -- that holds for arbitrary relations
of currents and deriving voltages in these antennas. We have proved analytically, numerically and experimentally that this decoupling is feasible for resonant dipole antennas.
A drastic decrease of mutual coupling was obtained for the case when the distance between two antennas was much smaller than one tenth
of the operation wavelength. The only drawback of this decoupling is the shrink of the lossless matching band by an order of magnitude (and the similarly narrow band of decoupling).
If the signal band is not correspondingly narrow, this shrink may be harmful for the antenna efficiency. In this case, resistive elements may further significantly reduce the efficiency. This is an open question what factor higher reduces efficiency.
However, for some applications (e.g. for MRI array coils) even a narrow operational band (relative bandwidth of the order of 0.1\%) reported in the present paper may be sufficient.
In our next paper we will expand the study to the case when the number of decoupled antennas is more than $N=2$.

\section{Acknowledgement}

This work was supported by the Ministry of Education and Science of the Russian Federation (project No. 14.587.21.0041 with the unique identi fier
RFMEFI58717X0041) and European Unions Horizon 2020 research and innovation program under grant agreement No. 736937.

\end{document}